\documentclass{PoS}
\usepackage{color}
\definecolor{red}{cmyk}{0,1,1,0.4}

\title{Vacuum Stability and the Higgs Boson}

\ShortTitle{Vacuum stability and the Higgs Boson}

\author{Jos\'e Ram\'on ESPINOSA
\\        
ICREA, Instituci\'o Catalana de Recerca i Estudis Avan\c{c}ats, Barcelona, Spain\\
IFAE, Universitat Aut{\`o}noma de Barcelona,
   08193~Bellaterra,~Barcelona, Spain\\
        E-mail: \email{espinosa@ifae.es}}

\abstract{The discovery of the Higgs boson at the LHC, and especially the determination of its mass around 125 GeV,  together with the absence of any trace of new physics make it conceivable that we live in a metastable (but long-lived) electroweak vacuum. I will describe the state-of-the-art calculation that leads to this conclusion, elaborate on possible implications as well as cures of this instability of the Higgs potential and discuss some possible lines of attack for lattice studies of such metastability. }

\FullConference{31st International Symposium on Lattice Field Theory LATTICE 2013\\
                 July 29 - August 3, 2013\\
                 Mainz, Germany}

\begin{document}

\section{Extrapolating the SM to Very High Scales and the Higgs Potential Instability}

The main result of the first run of the LHC was the discovery of the Higgs boson, with mass $M_H\simeq 126$ GeV \cite{higgsdiscovery}, which further study has 
shown to be compatible with the properties expected for a Standard Model (SM) Higgs, although there is still room for some deviation in its properties \cite{higgscouplings}. Besides this great success, no trace of physics beyond the SM (BSM) has been found, and this typically translates into bounds on the mass scale of different BSM scenarios, supersymmetric or otherwise,
of order the TeV \cite{LHCBSM}. If one is willing to hold on to the paradigm of naturalness, the hierarchy problem that afflicts the breaking of the electroweak (EW) symmetry would imply that BSM physics should be around the corner, probably on the reach of the LHC.
In this talk I take a different attitude: I disregard naturalness as a requisite for the physics associated to the breaking of the EW symmetry and I explore the possibility that the scale of new physics, $\Lambda$, could be as large as the Planck scale, $M_{Pl}$.

\begin{figure}[b]
$$\includegraphics[width=0.47\textwidth]{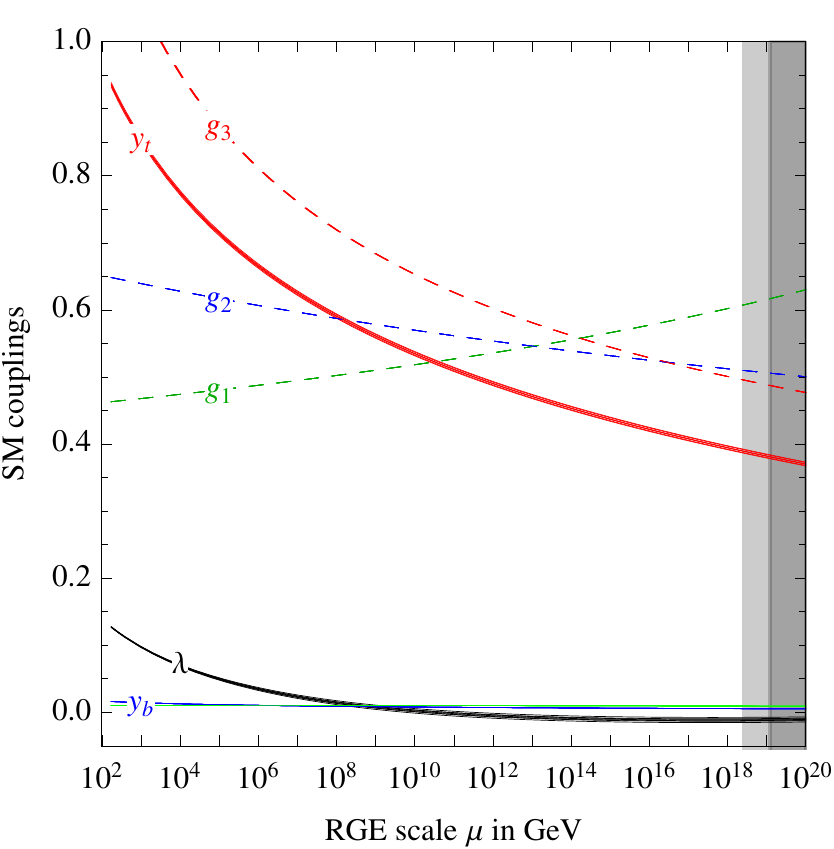}    \qquad
\includegraphics[width=0.49\textwidth]{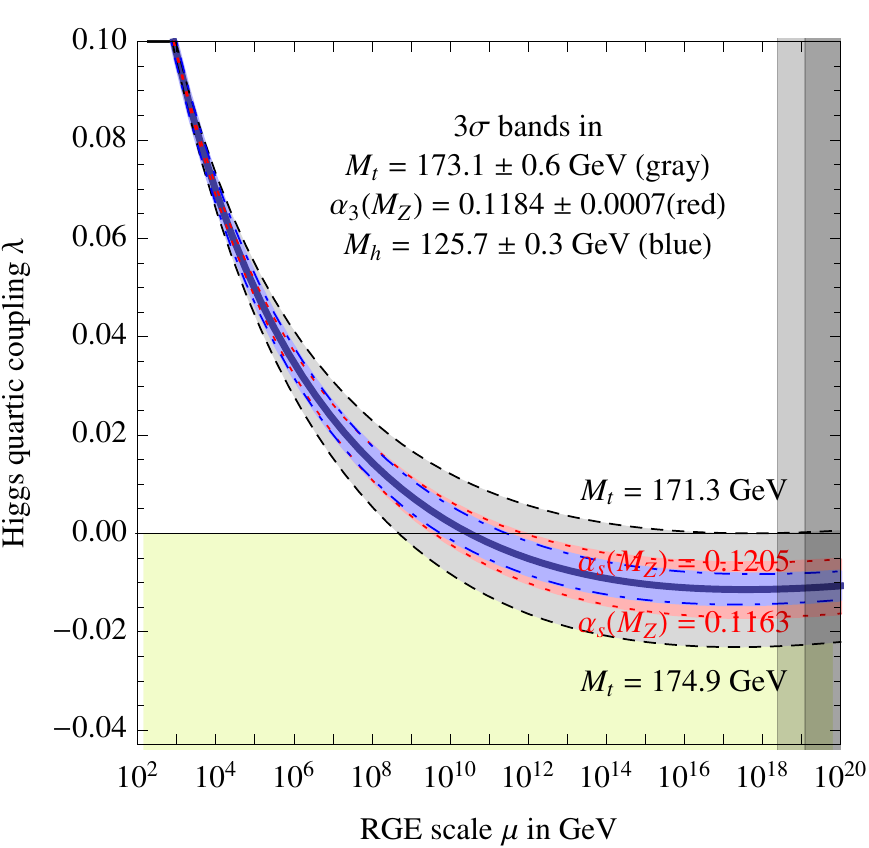}$$
\begin{center}
\caption{\label{fig:run}\emph{Left: Evolution of SM couplings from the EW scale to $M_{Pl}$. Right: Zoom on the  evolution of the Higgs quartic, $\lambda(\mu)$, for $M_h=125.7$ GeV, with uncertainties in the top mass, $\alpha_s$ and $M_h$ as indicated. (Plots taken from \cite{us}).}}
 \end{center}
\end{figure}

From that perspective, we have now in our hands a quantum field theory, the SM, that should then describe physics in the huge range from $M_W$ to $M_{Pl}$. All the model parameters have been determined experimentally, the last of them being the Higgs quartic coupling, fixed in this model by our knowledge of the Higgs mass.  Fig.~\ref{fig:run}, left plot, shows the running of the most important SM couplings extrapolated to very high energy scales using renormalization group (RG) techniques.
It shows the three $SU(3)_C\times SU(2)_L\times U(1)_Y$ gauge couplings getting closer in the ultraviolet (UV) but failing to unify precisely. It also shows how the top Yukawa coupling gets weaker in the UV (due to $\alpha_s$ effects, see below). The Higgs quartic coupling is also shown: it starts small at the EW scale, $\lambda(M_t)\sim 1/8$,
because the Higgs turned out to be light, and gets even smaller at
higher scales. The zoomed right plot in fig.~\ref{fig:run} shows that, in fact, $\lambda$ does something interesting: it gets negative at around $\mu\sim 10^{10}$~GeV.  

This steep behavior of $\lambda$ is due to the effect of one-loop top  corrections, which represent the dominant contribution to the beta function of $\lambda$, which describes the evolution of $\lambda$ with scale. One has, at one loop: 
\begin{equation}
\beta_\lambda =\frac{d\lambda}{d\log \mu}=
\frac{1}{16\pi^2}\left\{{\color{red}\bf
-6y_t^4 }+ 12y^2_t\lambda +
\frac{3}{8}\left[2g^4 + (g^2 + {g'}^2)^2\right]
-3\lambda (3g^2 + {g'}^2) + 24\lambda^2\right\}\ ,
\end{equation}
where $\mu$ is the renormalization scale and
$y_t$ is the sizable top Yukawa coupling, while $g$ and $g'$ are the $SU(2)_L$ and $U(1)_Y$ gauge couplings, respectively. Due to the dependence of $\beta_\lambda$ on the fourth power of $y_t$ (term in red) there is a strong dependence of the running of $\lambda$ on the top quark mass, as shown by the gray band in fig.~\ref{fig:run} (right), which corresponds to a $3\sigma$ variation of $M_t$ around is central value (as indicated). The bigger (smaller) $M_t$ is, the steeper (milder) the slope of the running $\lambda$. 

There is a smaller dependence of the running of $\lambda$ on the value of $\alpha_s$, which 
affects $\beta_\lambda$ indirectly through its effect on the running of $y_t$:
\begin{equation}
\beta_{y_t}=\frac{dy_t}{d\log\mu}=\frac{y_t}{16\pi^2}\left[
\frac{9}{2}y_t^2{\color{red}\bf -8g_s^2}-\frac{9}{4}g^2-\frac{17}{12}{g'}^2
\right]\ ,
\label{betayt}
\end{equation}
where $g_s$ is the $SU(3)_C$ gauge coupling.
This smaller effect is illustrated in the same fig.~\ref{fig:run} (right) by the thinner $3\sigma$ pink band, with higher (lower) $\alpha_s$ corresponding to softer (steeper) running. Finally, the thinnest band, in blue, corresponds to $3\sigma$ variations in the Higgs mass, as indicated. One also sees that $\lambda$ flattens out after getting negative: in that range of scales, gauge couplings become comparable in size with $y_t$ (see Fig.~\ref{fig:run}, left) and there is a cancellation leading to $\beta_\lambda\simeq 0$. As we will discuss in the last section, at even higher scales gauge couplings dominate, turning $\beta_\lambda$ positive and eventually making $\lambda>0$ (although this might happen beyond $M_{Pl}$).

The trouble with $\lambda$ becoming negative is that it will cause an instability in the Higgs potential. This is clear when one notices that:
{\em 1)} the potential at very high values of the field is dominated by the quartic term, and {\em 2)} a good approximation to the full potential at some field value $h$ requires that couplings are evaluated at a renormalization scale $\mu\sim h$ (see discussion in section 3). Therefore, at very high values of the Higgs field, the potential is $V(h\gg M_t) \simeq (1/4)\lambda(\mu=h) h^4$, which for $\lambda(h)<0$ is much deeper than our vacuum at the EW scale. This instability phenomenon, caused by heavy fermions coupled to light scalars, has been known since a long time ago \cite{oldies} and has been investigated since then in the SM with increasing degree of precision \cite{stability}, especially recently \cite{us0,lindner,shap,us,Buttazzo}, after it became apparent that the Higgs mass would lie in a very special region concerning the stability of the potential.

With the current precision in the Higgs mass determination and theoretical calculation of the stability bound (which will be reviewed in Sect.~3), one concludes that (given our theoretical assumptions about the absence of BSM physics) our vacuum would most likely be metastable.  We should then worry about its lifetime against decay through quantum tunneling to a deeper minimum at very high field values.

\section{Lifetime of the Metastable Electroweak Vacuum}

The decay probability rate of the EW vacuum per unit time and unit volume can be calculated by semiclassical methods \cite{tunnel}: it is basically $\sim h_I^4 \exp(-S_4)$, where $h_I$ is the value of the field around the region of instability (the only relevant mass scale in the problem), and $S_4$ is
the action of the 4-d Euclidean tunneling bounce solution, interpolating between the new phase at high field values and the EW phase.
A simple analytical approximation  for $S_4$  that captures the main effect\footnote{The tunneling rate has been calculated beyond tree level, including the effect of fluctuations around the bounce solution  in \cite{rateloop}. Gravitational effects, which have a negligible impact on the rate, were included in \cite{rategrav}.} is in fact possible for a negative-quartic potential $V\simeq -|\lambda(h)|h^4/4$: it is $S_4\simeq -8\pi^2/(3|\lambda(h_I)|)$,
showing the usual nonperturbative dependence on the coupling constant. The logarithmic dependence of $\lambda(h)$ on its
argument breaks the scale invariance of the classical potential 
and the tunneling occurs preferentially towards the scale $h_I$ for which $\lambda(h)$ takes its minimum value (or, what is the same, for which $\beta_\lambda=0$). 

\begin{figure}[b]
$$\includegraphics[width=0.47\textwidth]{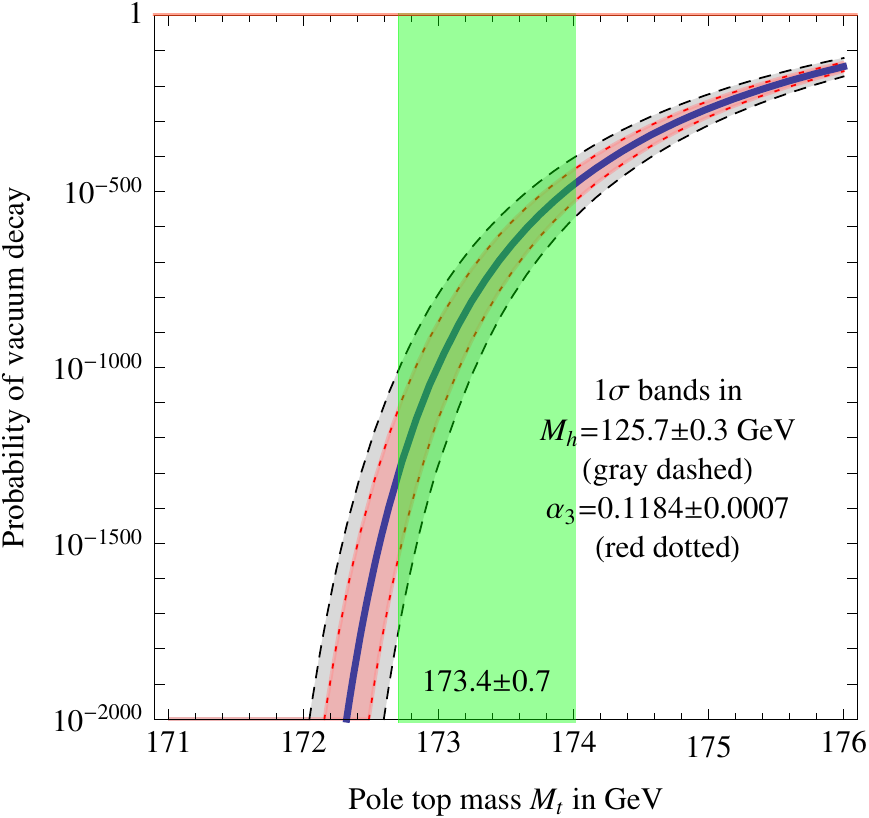}\qquad
\includegraphics[width=0.47\textwidth]{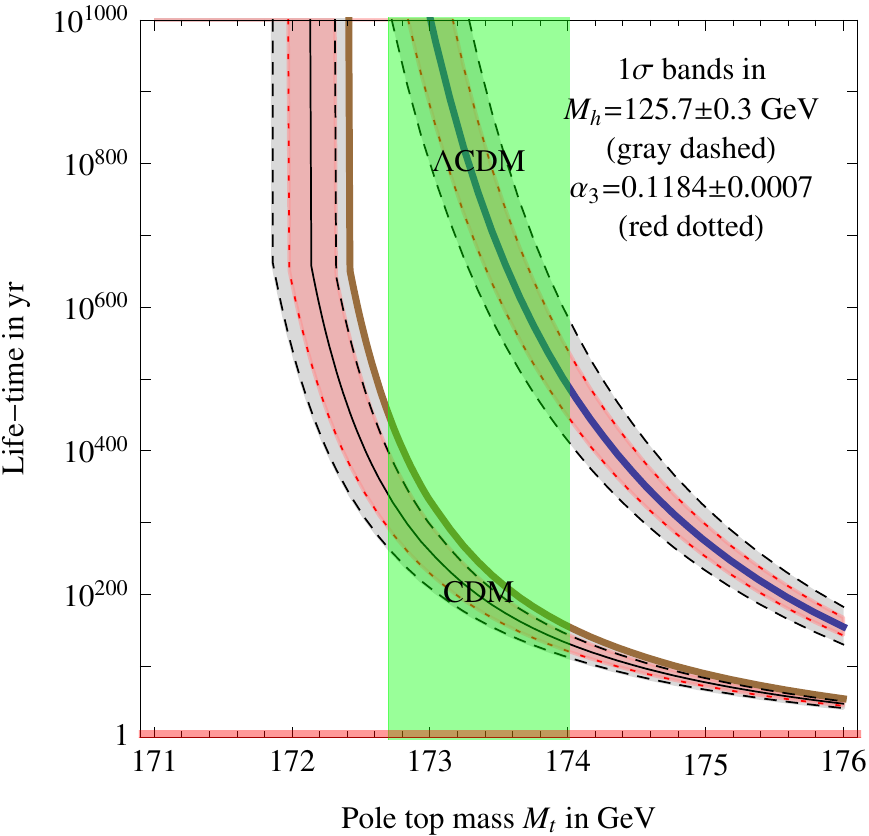}
$$
\begin{center}
\caption{\label{fig:life}\emph{Left: probability of EW vacuum decay by quantum tunneling as a function of $M_t$, for the measured values of $M_h$ and $\alpha_s$. Right: same, for the vacuum lifetime. The two branches correspond to different assumptions on the future of the universe evolution:
matter dominated (labeled CDM) or cosmological constant dominated (labelled $\Lambda$CDM). (Plots taken from \cite{Buttazzo}).}}
 \end{center}
\end{figure}

One gets for the decay rate the numerical estimate $dp/(dV \ dt)\sim h_I^4 exp[-2600/(|\lambda|/0.01)]$. This has to be multiplied by the 4-d space-time volume inside our past light-cone, which is basically the fourth power of the age of the Universe: $\sim \tau_U^4\sim (e^{140}/M_{Pl})^4$. It is then clear that the exponential suppression of the decay rate [for the observed value of $M_h$, which gives $\lambda(h_I)\sim -0.01$] wins over the large 
4-volume factor: the decay probability is extremely small, $p\ll 1$ or, in other words, the lifetime of the metastable EW vacuum $\tau_{EW}$ is extremely long, much larger than the age of the Universe, see Fig.~\ref{fig:life}.

\begin{figure}
$$\includegraphics[width=0.5\textwidth]{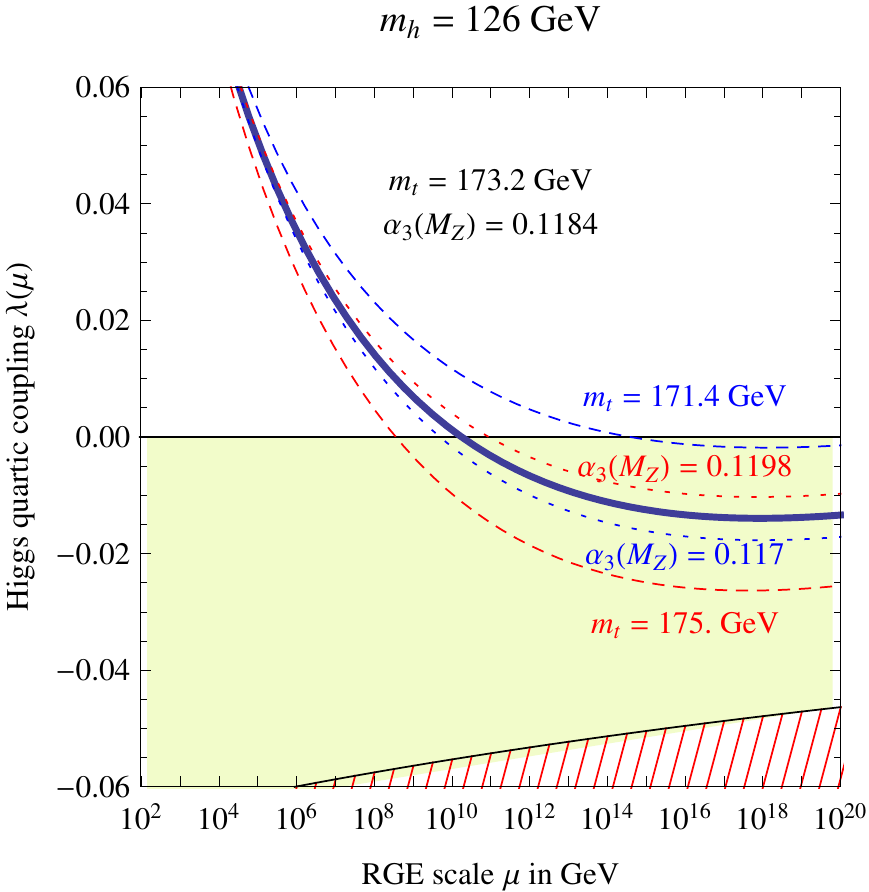}$$
\begin{center}
\caption{\label{fig:dangerlambda}\emph{Running Higgs quartic coupling, for $M_h=126$ GeV, showing the dependence on $M_t$ and $\alpha_s$ and the parameter region (red hatched) corresponding to an unstable vacuum with lifetime shorter than the age of the Universe. (Plot taken from \cite{us}).}}
 \end{center}
\end{figure}

This conclusion could have been different if the Higgs mass were smaller, resulting in a stronger instability of the Higgs potential. This point is illustrated by fig.~\ref{fig:dangerlambda}, which shows the (hatched) region of negative $\lambda(\mu)$ that would give a decay probability of the EW vacuum of order 1 or higher ({\it i.e.} a vacuum lifetime $\tau_{EW}$ smaller than $\tau_U$). We will call this region of parameter space the instability region. 

Besides the danger of vacuum decay by quantum tunneling, the EW vacuum could have also decayed in the early Universe by thermal excitations over the barrier that separates it from the deeper region of the potential at high field values \cite{thermaldecay}. The condition that the EW vacuum does not decay during those high-temperature stages of the early Universe (taken into account that the potential itself
is modified at finite temperature) can be used to set an upper bound on the reheating temperature $T_{RH}$ after inflation. However, it turns out that
for the current values of the Higgs mass, the potential is safe in such thermal environment and the bound on $T_{RH}$ would require lower
Higgs masses, $M_h\leq 122$ GeV, see \cite{us0}. Other cosmological implications can be found in \cite{cosmo}.

\section{NNLO Stability Bound and Implications}

From fig.~\ref{fig:dangerlambda} we also see that the possibility that the EW vacuum is absolutely stable up to $M_{Pl}$ would require values of $M_t$  and $\alpha_s$ in some $\sim 2-3\sigma$ tension with their central experimental values.
Traditionally, this possibility has been phrased in terms of the so-called stability bound on $M_h$: that is, how heavy should $M_h$ be to ensure a stable potential up to $M_{Pl}$.

The state-of-the-art determination of that stability bound \cite{us,Buttazzo} (at NNLO, see discussion below)  gives
\begin{equation}
M_h[GeV] > 129.6 + 2.0\, [M_t(GeV)-173.35]-0.5\left[ \frac{\alpha_s(M_z)-0.1184}{0.0007}\right]\pm 0.3\ .
\label{stabound}
\end{equation} 
In this expression, the main error comes from the experimental uncertainty in the top mass. From a naive combination of the experimental measurements from Tevatron and ATLAS plus CMS at the LHC one gets
$M_t=173.36 \pm 0.65_{exp} \pm 0.3_{th}$ GeV, see \cite{Buttazzo}.
The total 1$\sigma$ error for $M_t$ in Eq.~(\ref{stabound}) has been rounded up to 1 GeV to allow for a somewhat larger theoretical error, see the discussion below. Next in importance comes the error associated with the uncertainty in $\alpha_s(M_z)=0.1184\pm 0.0007$
\cite{alphas}. Finally, the last error is theoretical and comes from an estimate of higher order corrections, beyond NNLO. Such small error has been achieved only quite recently, with refs.~\cite{shap,us,Buttazzo} being the main contributors towards this goal. 

In order to achieve this precision one has to calculate reliably the
scalar potential in a wide range of field values, from the EW scale up to $M_{Pl}$.  There are potentially large logs, $\log[h/M_t]$, that need to be resummed and this can be done using standard renormalization group techniques. We expect that the $n^{th}$-loop contribution 
to the effective potential, $V_n$, will have log-enhanced terms
$\propto \alpha^n[\log(h/\mu)]^{n-k}$ [where $\alpha$ represents some perturbative expansion parameter, like $y_t^2/(16\pi^2)$], with a hierarchical ordering: the dominant leading-log order (LO) for $k=0$, next-to-leading-log order
(NLO) for $k=1$ and so on, till the $k=n$ non-log terms. Resummation
of these large logs is done by using the so-called RG-improved potential: a potential with couplings (and field) that are running with the renormalization scale, see {\it e.g.} \cite{RGV}, which is then chosen as $\mu\sim h$. In this way, a tree-level expression $V_0$ for the potential, with couplings running with their 1-loop beta functions resums the LO terms {\it to all loops}. A one-loop
expression for the potential, with couplings running with their two-loop 
RG equations resums also the NLO terms {\it to all loops}, and so on. To match a given level of precision, the relations between running couplings and the observables that determine them has to be performed at the corresponding level of accuracy: tree-level matching for LO (as a loop-order error in the matching propagates to NLO corrections only), one-loop matching for NLO, etc. The ingredients for the NNLO calculation of the stability bound are then clear: use the RG-improved two-loop effective potential \cite{V2}, in which couplings are running with 3-loop beta functions \cite{beta3} and use 2-loop matching \cite{shap,us,Buttazzo} to relate $\lambda$ and $y_t$ to $M_h$ and $M_t$.

\begin{figure}
$$\includegraphics[width=0.69\textwidth,height=0.4\textwidth]{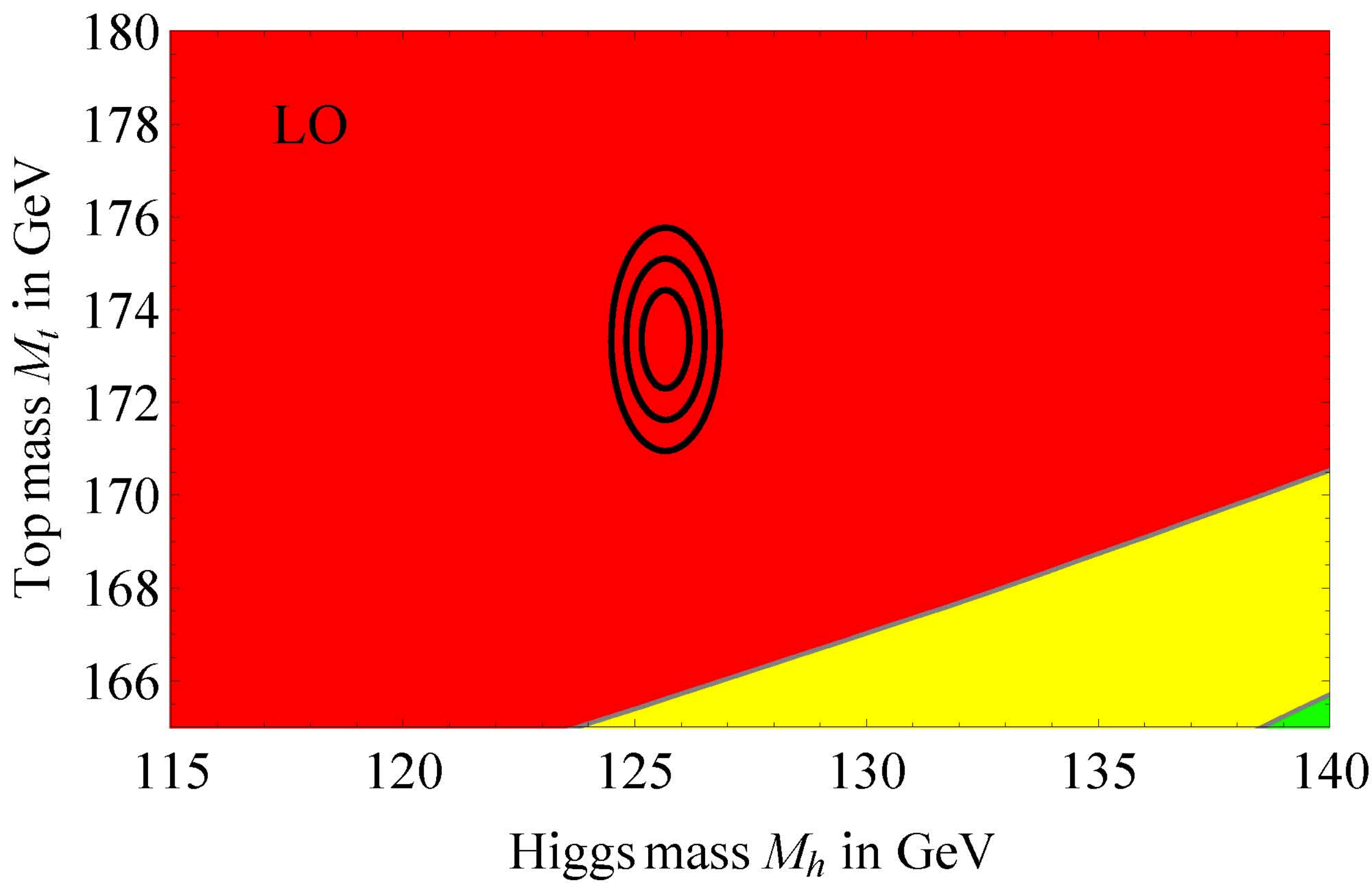}    
$$
$$
\includegraphics[width=0.69\textwidth,height=0.4\textwidth]{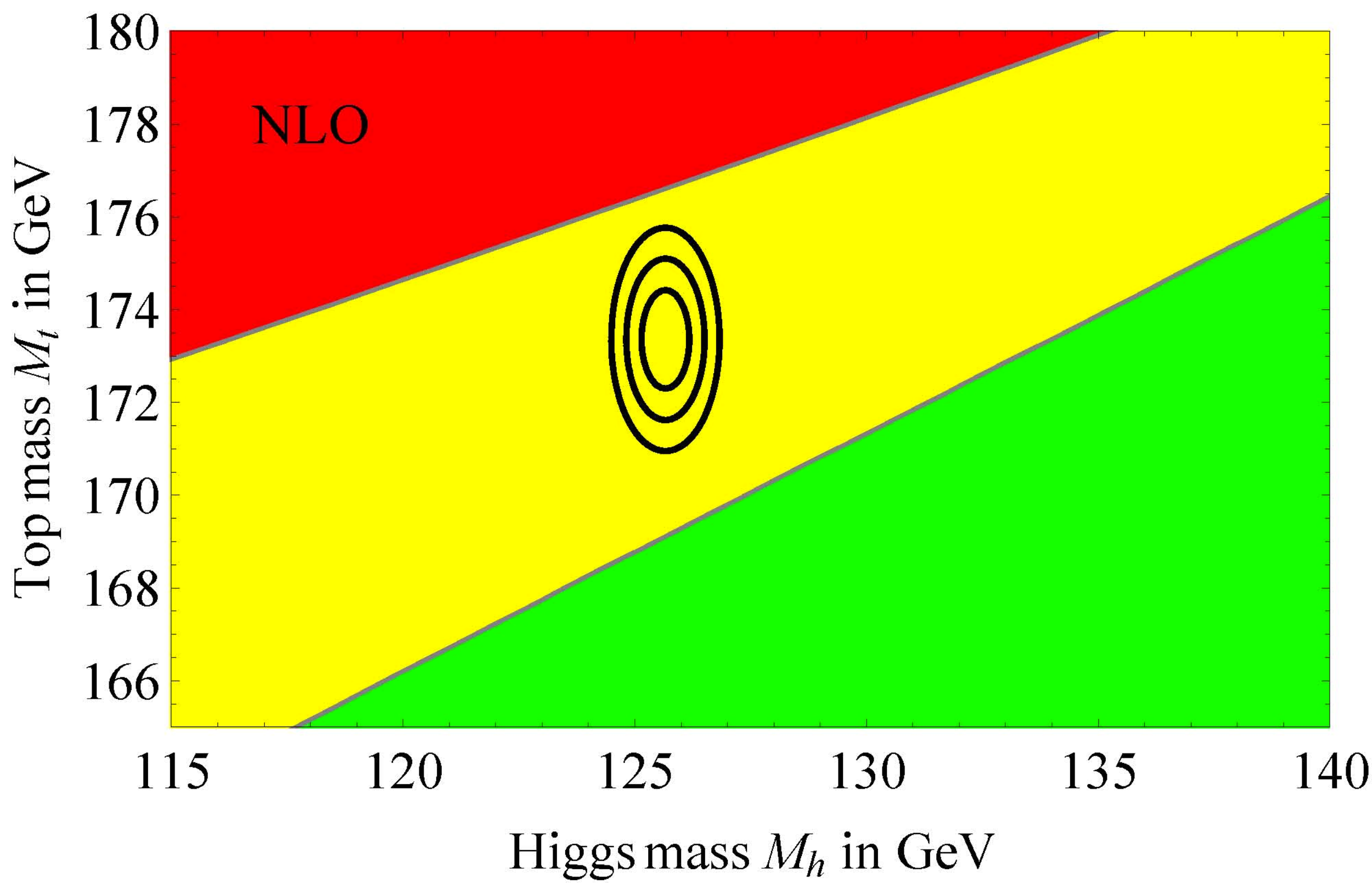}  $$
$$ 
\includegraphics[width=0.69\textwidth,height=0.4\textwidth]{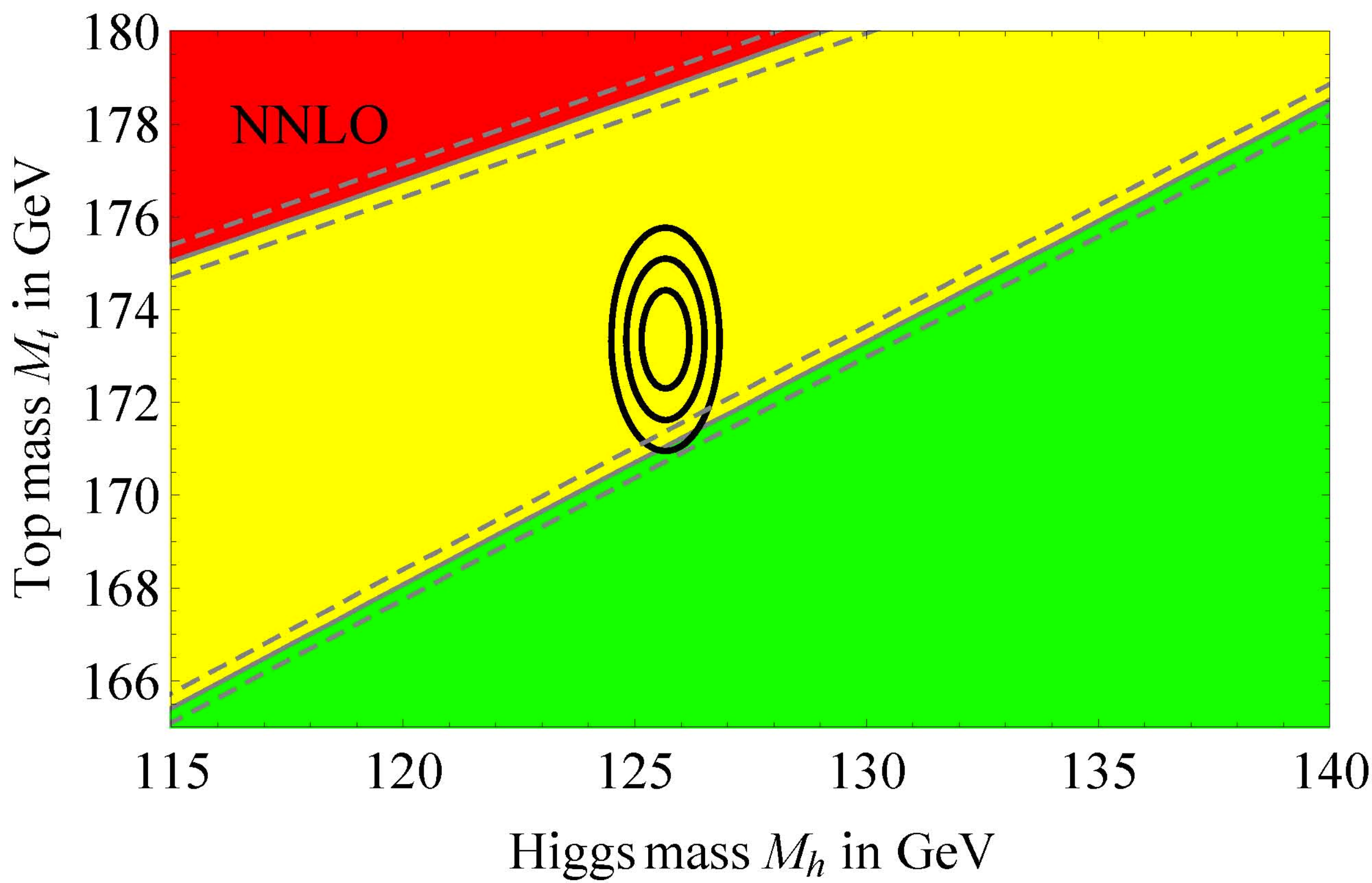} 
$$
\begin{center}
\caption{\label{fig:progress}\emph{Regions in the $(M_h,M_t)$ parameter space corresponding to absolute stability (green), metastability with lifetime $\tau_{EW}$ longer than $\tau_U$ (yellow), and
instability, with $\tau_{EW}<\tau_U$, of the EW vacuum. The ellipses give the experimental values at 1, 2 and $3\sigma$. The different versions correspond to progressively more precise calculations, from LO to NNLO as indicated. }}
 \end{center}
\end{figure}

\begin{figure}[t]
$$ 
\includegraphics[width=0.45\textwidth,height=0.45\textwidth]{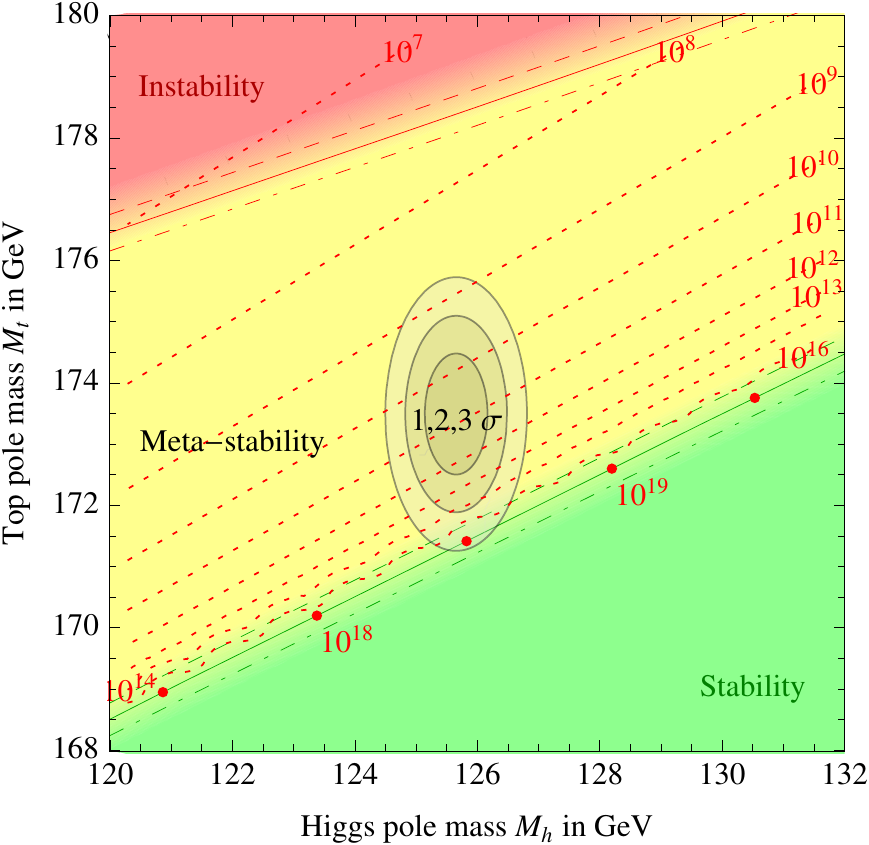} \qquad
\includegraphics[width=0.44\textwidth,height=0.44\textwidth]{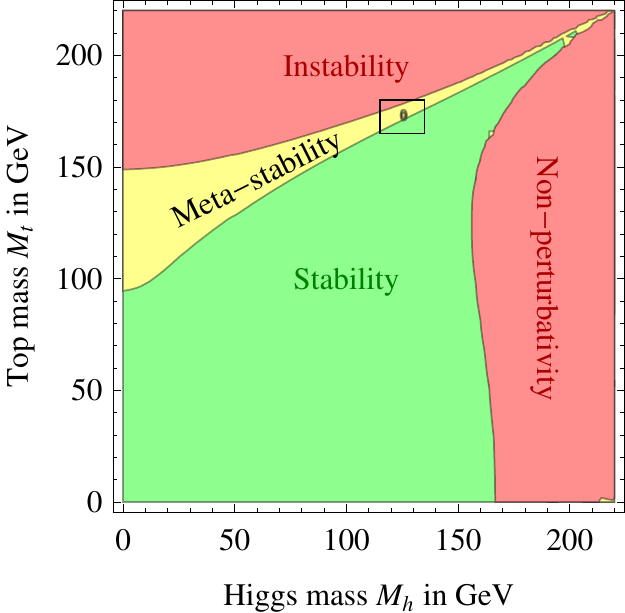}
$$
\begin{center}
\caption{\label{fig:edgy}\emph{Regions in the $(M_h,M_t)$ parameter space corresponding to absolute stability (green), metastability with lifetime $\tau_{EW}$ longer than $\tau_U$ (yellow), and
instability, with $\tau_{EW}<\tau_U$, calculated at NNLO. 
The ellipses give the experimental values at 1, 2 and 3 $\sigma$. 
The red-dashed lines in the zoomed-in version (left, from \cite{Buttazzo}) indicate the scale of instability, in GeV. The zoomed-out version (right, from \cite{us}) also shows the region corresponding to non-perturbative Higgs quartic below $M_{Pl}$. }}
 \end{center}
\end{figure} 
In order to illustrate the need of such very precise calculation of this stability bound to determine the properties of the EW vacuum, given our precise knowledge of $M_h$ and $M_t$, fig.~\ref{fig:progress} shows the regions in $(M_h,M_t)$ parameter space corresponding to an EW vacuum that is stable (green area), metastable (with lifetime $\tau_{EW}$ longer than $\tau_U$, yellow area) or unstable (with  $\tau_{EW}<\tau_U$, red region). The plots in Fig.~\ref{fig:progress} show the location of these regions resulting from a LO, NLO and NNLO calculation, from top to down. The experimental ellipses for $M_h$ and $M_t$ are also shown.

This figure demonstrates that NNLO precision is crucial to answer questions about the stability of the EW vacuum. What about higher order (NNNLO) corrections? The NNLO plot shows also (dashed lines) the remaining error, obtained by combining in quadrature the (rather small) theoretical error expected from the non-inclusion of such higher order corrections and the uncertainty from $\alpha_s$: clearly a definitive answer to the stability question will require a better knowledge of the top mass rather than an even more refined theoretical calculation. 
In terms of the top mass, the stability bound reads \cite{Buttazzo}:
\begin{equation}
M_t < (171.36\pm  0.15 \pm 0.25_{\alpha_s} \pm 0.17_{M_h})\, {\mathrm GeV} = (171.36 \pm 0.46)\, {\mathrm GeV}\ ,
\end{equation}
where, in the last expression, the theoretical error is combined in quadrature with the indicated experimental uncertainties from $\alpha_s$ and $M_h$.

Concerning the impact of $M_t$ on the stability bound, there is some controversy in the literature regarding the relationship between the top mass measured at the Tevatron and LHC and the top pole-mass. Although the naive expectation would assign an error of order $\Lambda_{QCD}$ to the connection between these two numbers, a more drastic proposal has been advocated in \cite{mt}: to use instead the running top mass measured through the total production cross-section $\sigma(pp/p\bar p\rightarrow t\bar{t}+X)$ at Tevatron and the LHC, which allows for a theoretically cleaner determination of $M_t$. However, this leads to a value of the top mass compatible with the Tevatron and LHC values but with an error which is a factor of 4 worst: $M_t=173.3\pm 2.8$ GeV \cite{mt}. Of course, if one is willing to downgrade the error on $M_t$ in this way, there would still be room for absolute stability up to $M_{Pl}$ by moving into the lower range for $M_t$. Clearly,  a better understanding of the theoretical errors
in the top mass determination would be desirable. See \cite{mtreview}
for a review of the issues involved, current status and future expectations (presumably down to $\delta M_t\sim 500-600$ MeV at the LHC) concerning this important measurement. 

\begin{figure}[t]
$$ 
\includegraphics[width=0.49\textwidth,height=0.49\textwidth]{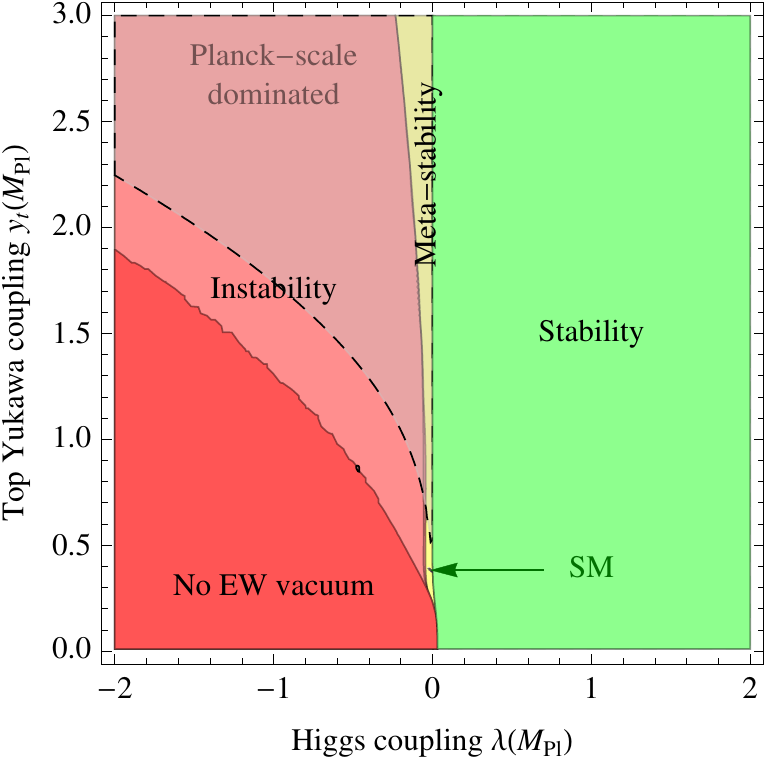}
$$
\begin{center}
\caption{\label{fig:edgyPl}\emph{Different regions for the Higgs potential behaviour in the $\{\lambda(M_{Pl}),y_t(M_{Pl})\}$ plane. }}
 \end{center}
\end{figure}

In Fig.~\ref{fig:edgy}, the left plot shows again the different regions
concerning stability of the EW vacuum calculated at NNLO, with further information on the scale of instability, in red dashed lines. The right plot shows the same NNLO stability regions [plus the region in which $\lambda(\mu)$ becomes non-perturbative below $M_{Pl}$] in a zoomed-out range for Higgs and top masses. This last version emphasizes the fact that we seem to be living in a very untypical region of parameter space, really close to the boundary for absolute stability in the narrow wedge for a long-lived EW vacuum. A complementary view of the same observation is offered by Fig.~\ref{fig:edgyPl}, which plots the different regions for the Higgs potential behaviour in the $\{\lambda(M_{Pl}),y_t(M_{Pl})\}$ plane (as these parameters should be more fundamental). There is a new phase
without a vacuum at the EW scale for $\lambda(M_{Pl})<0$ and small values of $y_t(M_{Pl})$ and, inside the instability region, the dashed line delimits the range for which Planck-scale physics can play a significant role in determining the high-scale behavior of the potential.
The SM location in the narrow metastability wedge is indicated by an arrow, showing once again how atypical our universe looks like.

This intriguing fact has triggered many speculations concerning its possible significance \cite{shap,us, Buttazzo} including: high-scale Supersymmetry \cite{HighSUSY}, enforcing $\lambda(\Lambda)=0$ through $\tan\beta=1$; IR fixed points of some asymptotically safe gravity \cite{IRgrav},
among other ideas (even some that predate the Higgs discovery \cite{MPP}). Is $\lambda(M_{Pl})\simeq 0$ related to the fact that we also live very close to a second phase boundary, the one separating the EW broken and unbroken phases? This boundary is associated to the fact that the mass parameter in the Higgs potential, $m^2$, is extremely small in Planck units: $m^2/M_{Pl}^2\sim 0$. In this respect, it seems that 
the Higgs potential has a very particular form at the Planck scale, with both $\lambda$ and $m^2$ being very small. In addition, also $\beta_\lambda$ takes a special value $\simeq 0$ not far from $M_{Pl}$. Why do EW parameters seem to take such intriguing values at the Planck scale, the scale of gravitational physics, which is totally unrelated to the EW scale? No compelling theoretical explanation 
has been offered so far.

\section{Vacuum Instability and Physics Beyond the Standard Model}

Needless to say, the intriguing "near-criticality" discussed in the previous section could be just a mirage if new BSM physics appears below $M_{Pl}$ in such a way that the running of $\lambda(\mu)$ is modified significantly. Notice however, that the existence of the instability cannot be used by itself as a motivation for BSM, given the huge EW vacuum lifetime. Nevertheless, we do expect new physics BSM, {\it e.g.} to explain dark matter, neutrino masses or the matter-antimatter asymmetry and it is natural to ask how such physics could affect the near-criticality issue. 

We can distinguish three possibilities concerning the impact of  new physics on the stability of the Higgs potential: {\em a)} it can make the stability worse; {\em b)} be irrelevant; or {\em c)} cure it. 
It is easy to find examples of the three options, {\it e.g.} in the framework of see-saw neutrinos, say of type I.  In such scenario, neutrinos impact the running of $\lambda(\mu)$ through their Yukawa couplings,
which scale like $y_\nu^2 \sim M_N m_\nu/v^2$, where $m_\nu$ is the mass of the lightest neutrinos, $M_N$ the mass of the heavy right handed ones and $v=246$ GeV is the Higgs vacuum expectation value. 

\begin{description}
\item[a)] For sufficiently large $M_N$, the destabilizing effect of a large $y_\nu$ can make the instability much worse, even reducing the vacuum lifetime below $\tau_U$ [if $\lambda(\mu)$ is driven by this effect into the dangerous hatched region in Fig.~\ref{fig:dangerlambda}]. This would be in contradiction with our existence and can be used to set an upper bound on $M_N$, see \cite{casas,us0}.

\item[b)] For values of $M_N$ significantly smaller than this upper bound, of order $M_N\simeq 10^{13-14}$ GeV for $m_\nu\simeq 0.06-1$ eV, the neutrino Yukawas would be too small to have a significant effect on the running of $\lambda$ and their presence would be irrelevant for the potential instability discussed in previous sections. 

\item[c)] Finally,  a see-saw scenario that cures the instability is easy to build using a powerful stabilization mechanism through a heavy singlet field $S$ coupled to the Higgs as $\lambda_{HS} S^2 |H|^2$ and having a nonzero 
vacuum expectation value. When $S$ is decoupled, the low-energy $\lambda$ is reduced by a negative threshold effect.  The apparent instability of the potential is a mirage, as $\lambda$ above the $S$ threshold is larger than a naive extrapolation in the pure SM indicates. This mechanism can be made fully compatible with a see-saw mechanism in which $M_N$ is generated by the singlet vacuum expectation value, taken to be smaller than the SM instability scale $\sim 10^{10}$ GeV, and satisfying the lower constraints on $M_N$ from leptogenesis \cite{singlet}. 
\end{description}
Obviously, other stabilization mechanisms exist, and almost all extensions of the SM at the TeV scale will modify the behavior (or very existence) of the Higgs field at high energies. In any case, potential stability (at least in the weak sense of demanding $\tau_{EW}\gg\tau_{U}$) can be used to constrain BSM models that do not guarantee (unlike Supersymmetry) a good UV behavior of the Higgs potential.

As an example, if there is in fact an instability of the potential below the Planck scale, the minimal scenario of Higgs inflation \cite{HI} (which already is known to suffer from a unitarity/naturalness problem \cite{HItrouble}) cannot be realized \cite{us0,alb}: the mechanism claimed to give a plateau at high field values requires that the potential
grows like $\lambda h^4$ in the UV, with positive $\lambda$.
One is then lead to non-minimal options that must cure, not only the unitarity problem  but also the instability. This could in principle be achieved in the scenario proposed in  \cite{GL}, through the singlet stabilization mechanism discussed above, although the range of parameters required is somewhat contrived \cite{singlet}.

\section{Vacuum Instability in the Lattice}

One expects the perturbative continuum approach to the calculation of stability bounds to be reliable, as the couplings remain in the perturbative range. Nevertheless, stability bounds have also been studied in the lattice in models with scalars and fermions, expected to suffer
from this generic instability phenomenon. In the lattice, the stability bound appears as a lowest possible value for the scalar mass, value which is associated with the lowest possible bare scalar quartic coupling
[or $\lambda(\Lambda)=0$, where $\Lambda$
is the cutoff scale].

\begin{figure}[b]
$$\includegraphics[width=0.5\textwidth]{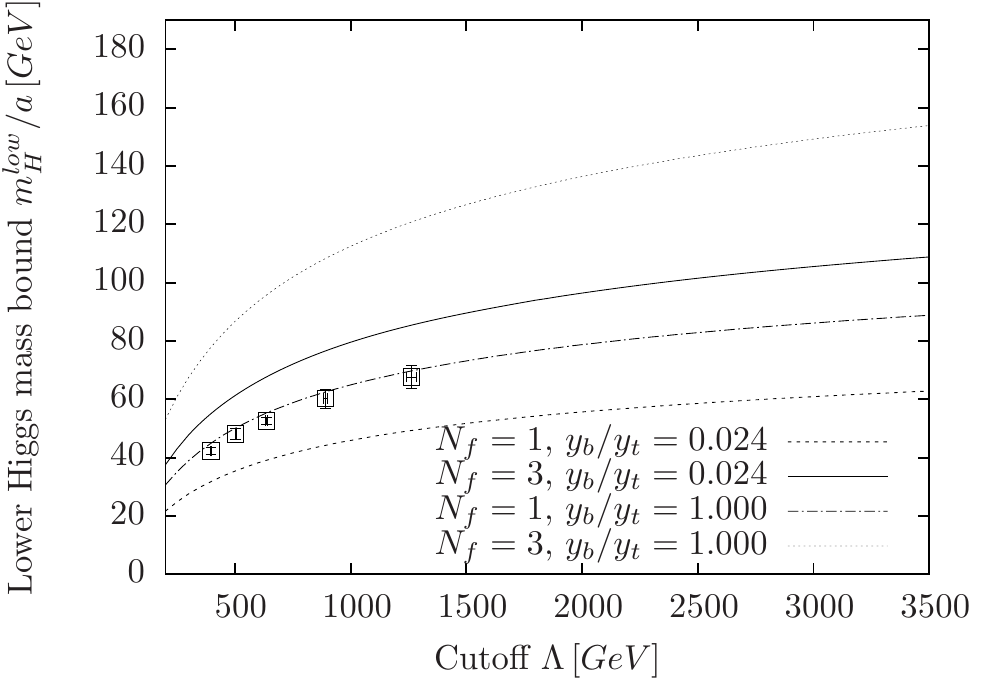}$$
\begin{center}
\caption{\label{fig:lattice}\emph{Lower stability bounds on $M_h$ from the lattice analysis of a Higgs-Yukawa model, taken  from \cite{Karl}.}}
 \end{center}
\end{figure}

The state-of-the-art lattice analyses of stability bounds derived such bounds in a chirally invariant Higgs-Yukawa model \cite{Karl} (see also \cite{Zoltan}), which should capture the main key ingredients of the phenomenon in the SM: a scalar Higgs doublet and a fermion $(t,b)$ doublet coupled through Yukawa interactions. Lattice perturbation theory is used to obtain a lower bound $M_h(\Lambda)$
associated with $\lambda(\Lambda)=0$, bound which is later confronted with lattice simulations done for the simplest case
with $N_f=1$ and $y_t=y_b$. After checking good agreement, the calculation in lattice perturbation theory is extrapolated to the more realistic case with $N_f=3$ and $y_t/y_b=0.024$. The bounds obtained are shown in Fig.~\ref{fig:lattice}, as a function of the cutoff scale. As expected, the bound is a bit higher (or the instability scale is significantly lower) than what one gets in the Standard Model for the same values of the Higgs mass, the reason being due to the non-inclusion of gauge couplings in the lattice analysis [as we saw, in the SM the top Yukawa coupling runs to smaller values in the UV due to $\alpha_s$, see Eq.~(\ref{betayt})]. 
From the continuum analysis we know that, if the effect of gauge couplings were included, the instability scale for $M_h\simeq 126$ GeV would appear at extremely large scales, inaccesible to lattice simulations.

Nevertheless, besides confirming that the stability bound is indeed there, lattice simulations could also be useful to study some truly non-perturbative effects associated with the physics of a metastable vacuum. In particular, they could be used to study the tunneling process by which the metastable vacuum decays. One (apparent) obstacle for this, which has been previously discussed in the literature, is the need of accessing the field range with negative values of the Higgs quartic coupling, as this is the field range at which the potential is lower than the EW vacuum, and towards which the tunneling occurs. This is a problem for the lattice as properly defining the theory being put in the lattice requires $\lambda(\Lambda)\geq 0$. This difficulty, which lead some authors to even doubt the very existence of an instability in the potential \cite{KutiHolland}, can be easily circumvented, as we will see next. 

First, this is not a problem for the continuum calculations: one simply assumes that some BSM physics will appear at some scale heavier than the instability scale, eventually stabilizing the potential (and therefore creating a new 
vacuum at large field values, which is deeper than the EW one). The actual calculation of the decay rate of the EW vacuum is not sensitive to the details of this stabilization under some reasonable assumptions: that it occurs well above the instability scale (which is roughly the scale that controls the tunneling and can be orders of magnitude below the scale at which the minimum appears) and that the instability of the potential is not made worst by the heavy new physics before it gets stabilized at even higher energies (a condition violated by the analysis in \cite{UVtrouble}).
 
\begin{figure}[t]
$$\includegraphics[width=0.45\textwidth,height=0.3\textwidth]{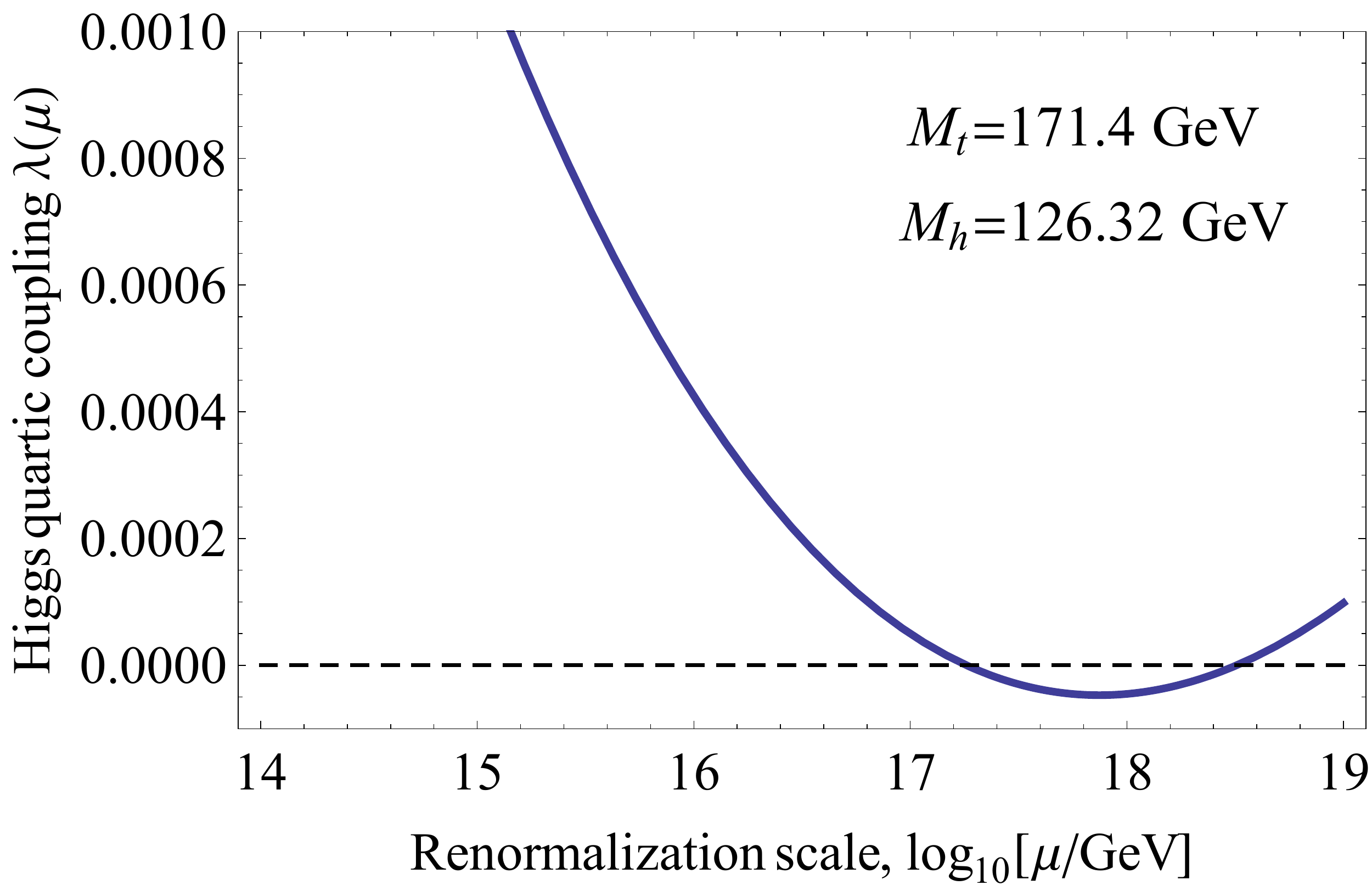}\qquad
\includegraphics[width=0.45\textwidth]{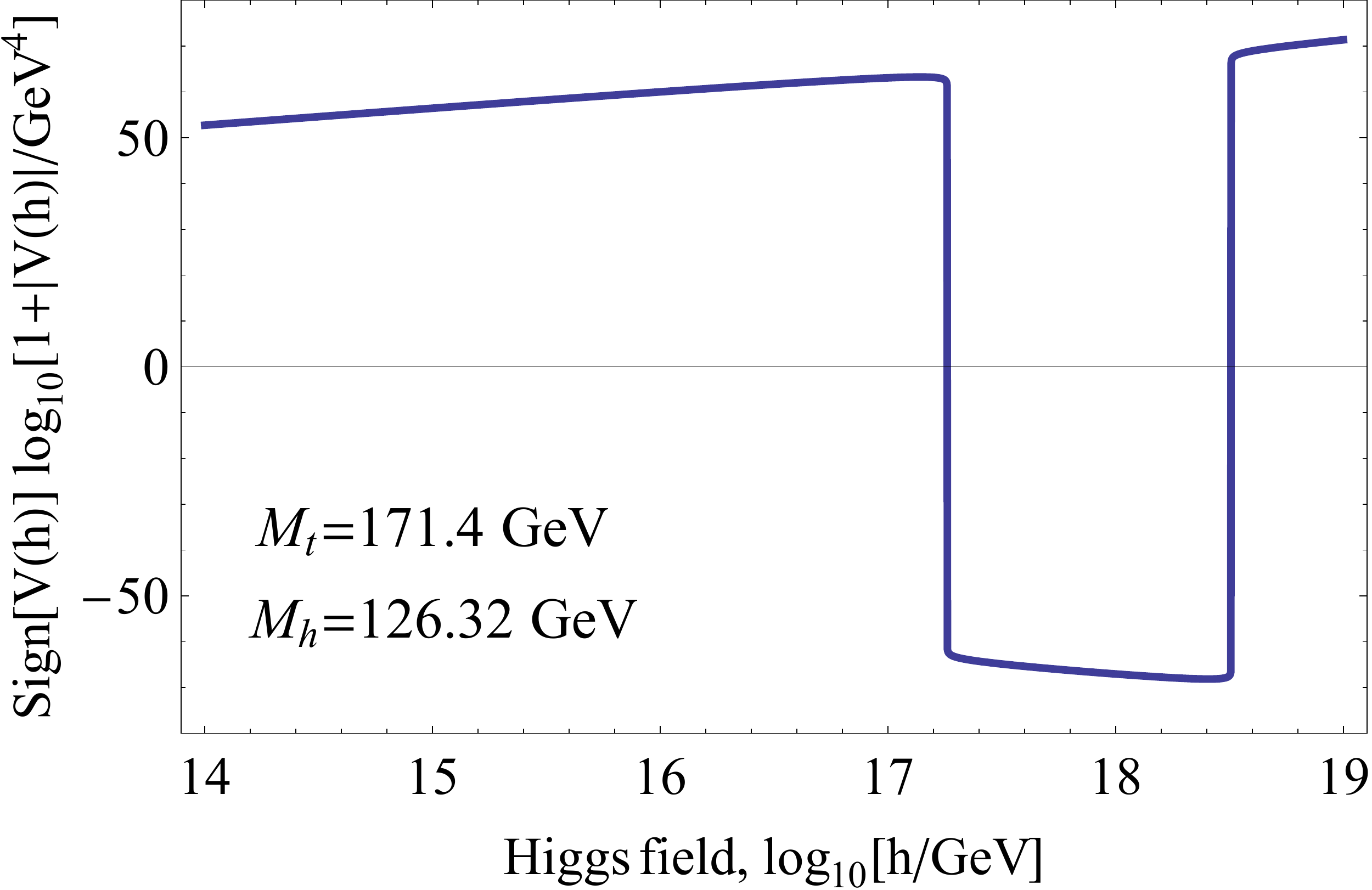}
$$
\begin{center}
\caption{\label{fig:dip}\emph{Example of parameter choice leading to a Higgs quartic coupling getting negative in a limited range of scales, with $\lambda(\Lambda)>0$ (left plot) and the corresponding Higgs potential (right plot), showing the deep non-standard minimum.}}
 \end{center}
\end{figure}

Moreover, in the pure SM there are some special choices of $M_t$, $M_h$ and $\alpha_s$ for which the unstable potential gets stabilized in the UV
simply by the RG flow of couplings: with $y_t$ getting smaller and smaller in the UV (due to the effect of $\alpha_s$), eventually EW gauge couplings make $\beta_\lambda$ positive and $\lambda(\mu)$ turns positive after an interval in which it is negative. This is a well known possibility, illustrated by Fig.~\ref{fig:dip}, which shows the running $\lambda(\mu)$ and 
the corresponding Higgs potential (in a log-log plot) for one such parameter choice. 
Such behavior demonstrates in principle that it should be possible to study the potential instability in the lattice: one can start at some heavy cutoff with a well defined theory with $\lambda(\Lambda)>0$,
in spite of which,
the theory can develop an instability at some intermediate range of scales that can in principle be much smaller than the cutoff (if one wishes to
get rid of potential cutoff artifacts).  
 
As we have mentioned already, relying on $\alpha_s$ as a way to diminish the destabilizing impact of the Yukawa coupling results in an instability scale that is many orders of magnitude above the electroweak scale, beyond the reach of lattice simulations. In order to be able to study in the lattice the decay of the metastable vacuum, some other (more efficient) stabilization mechanism should be used
to ensure that the true minimum is closer to the EW scale.
One possibility is to use the singlet mechanism \cite{singlet} discussed in the previous section, but in this case choosing the mass of the singlet somewhat above the instability scale, in such a way that stabilization of the potential occurs only after an interval of scales with $\lambda(\mu)<0$. Another possibility, which is currently under investigation (see talk by 
Attila Nagy \cite{attila} at this conference), is to stabilize the potential in the UV by higher order operators.

\section{Conclusions}

We finally have data to explore the physics of EW symmetry breaking.
So far, we have learned that the breaking is associated with a Higgs scalar of mass $M_h\simeq 126$ GeV that looks very much compatible with SM expectations, although there is still room for deviations in the properties of this scalar particle. On the other hand, the promise of nearby BSM physics, based on naturalness arguments, has not been
fulfilled yet. If one is willing to take this as indication of a fine-tuned
Higgs sector, extrapolation of the SM to very high energies reveals
a potentially dangerous instability in the Higgs potential. According to this, we would be living in a metastable EW vacuum, which however has a lifetime enormously large compared with the age of the Universe.

The previous  statement can be understood as resulting from the fact
that we live in a very particular region of parameter space in the  $(M_h,M_t)$ plane: very close to the boundary that separates the region of full stability of the potential from that of metastability.
Only time will tell whether there is a deeper meaning in that intriguing fact or whether new physics awaits us in the next LHC run (as expected from naturalness of EW breaking) that would expose this coincidence as a pure accident.

\section*{Acknowledgments}

I would like to thank my collaborators on this topic over many years:
M. Quir\'os, A. Casas, A. Riotto, G. Giudice, J. Ellis, A. Hoecker, A. Strumia, H.M. Lee, G. Isidori, J. Ellias-Mir\'o, G. Degrassi and S. di Vita.
This work has been partly supported by Spanish Consolider Ingenio 2010 Programme CPAN (CSD2007-00042) and the Spanish Ministry MICNN under grants FPA2010-17747 and FPA2011-25948; and the Generalitat de Catalunya grant 2009SGR894.

\end{document}